\documentclass[
    ,final            
  ,cmfonts
  ]
  {aipproc}

\layoutstyle{6x9}

\begin{document}

\title{A Model for the Generalized Parton Distribution of the Pion}


\author{P.Stassart}{
  address={Physique Th\'eorique Fondamentale, Bât. B5a, Univ. de Li\`ege, Sart Tilman, B-4000 Li\`ege 1}
}

\author{F.Bissey}{
  address={Institute of Fund. Sciences, Massey Univ., Private Bag 11 222, Palmerston North, NZ}
}

\author{J.R.Cudell}{
  address={Physique Th\'eorique Fondamentale, Bât. B5a, Univ. de Li\`ege, Sart Tilman, B-4000 Li\`ege 1} 
}

\author{J.Cugnon}{
  address={Physique Th\'eorique Fondamentale, Bât. B5a, Univ. de Li\`ege, Sart Tilman, B-4000 Li\`ege 1}
}

\author{J.P. Lansberg}{
  address={Physique Th\'eorique Fondamentale, Bât. B5a, Univ. de Li\`ege, Sart Tilman, B-4000 Li\`ege 1}
}

\begin{abstract}
We calculate the off-forward structure function of the pion
within a simple model where the size of the pion is accounted for using a
momentum cut-off. Twist-two and twist three generalized parton
distributions are extracted. Relations between twist-three and twist-two 
contributions are obtained, the origin of which is not kinematical as they differ
from those arising from the Wandzura-Wilczek approximation.
\end{abstract}

\maketitle


\section{Introduction}

Recent interest has focused on off-forward parton distributions \cite{off}
as they carry information on correlations between parton inside hadrons.
Based on the model we built to calculate the diagonal structure functions of
the pion in a gauge-invariant, regularization-independent way \cite{ours},
we have performed the calculations of off-forward structure functions 
and link them to generalized parton distributions.

In this simple model, the pion field is related to the constituent quarks
through a $\gamma^5$ vertex and pion size effects are introduced through  a
gauge-invariant cut-off procedure, by requiring that the relative momentum squared 
of the quarks inside the pion be smaller than the cut-off value.

In the following, we shall calculate the imaginary part of the off-forward 
scattering amplitude and link it to the structure functions that are the 
coefficients of the 
five independent tensors in this amplitude. We shall write these structure functions
in terms of vector and axial-vector form factors \cite{Bel}
and link them to the generalized parton distributions
${H}$, ${H}^3$ and $\tilde {H}^3$.

\section{The Model}

\begin{figure}[h]
\centering\mbox{\includegraphics[height=1.8cm]{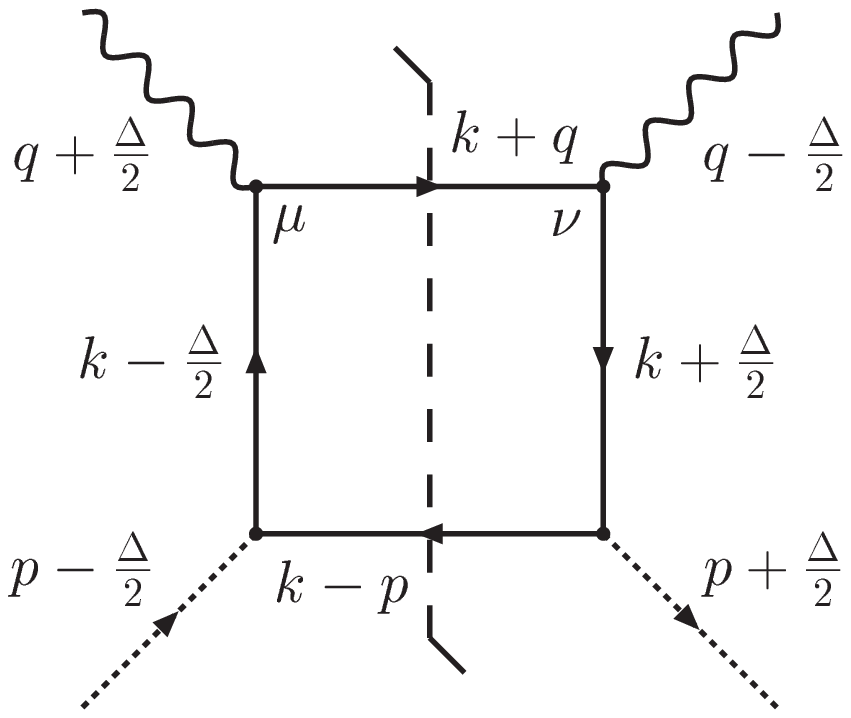}\quad  
\includegraphics[height=1.8cm]{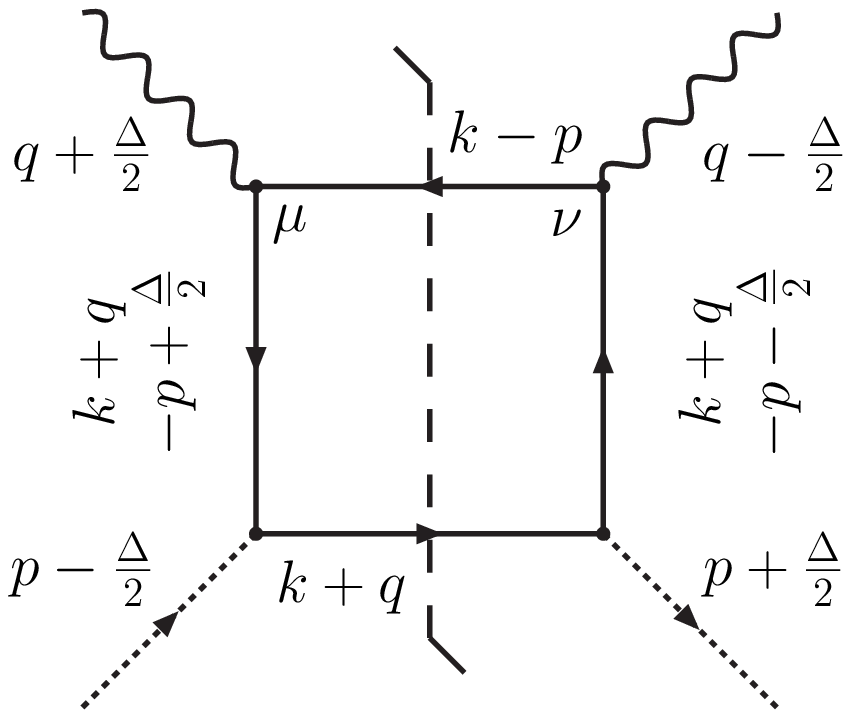}\quad
\includegraphics[height=1.8cm]{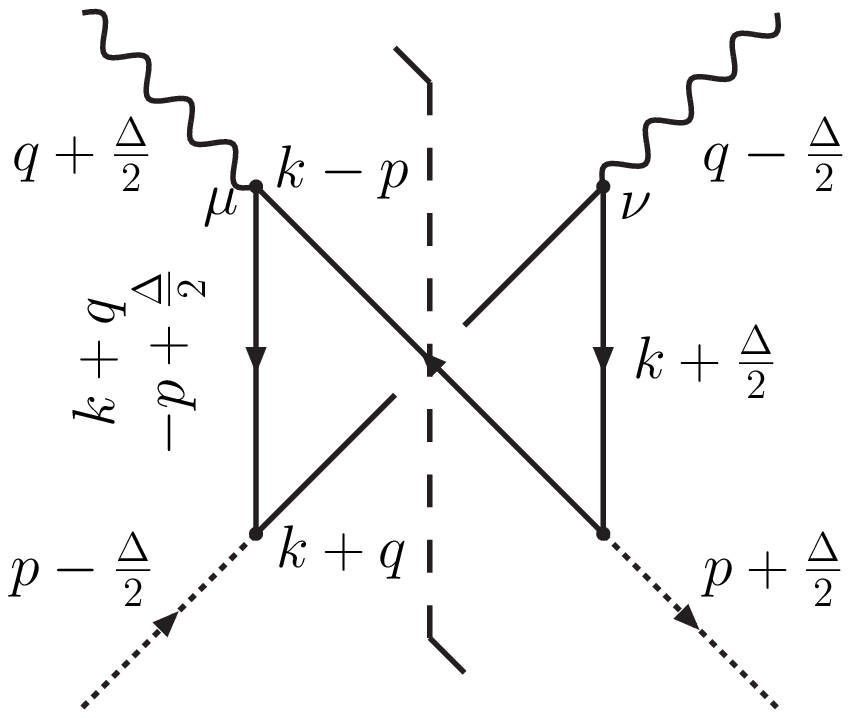}\quad  \includegraphics[height=1.8cm]{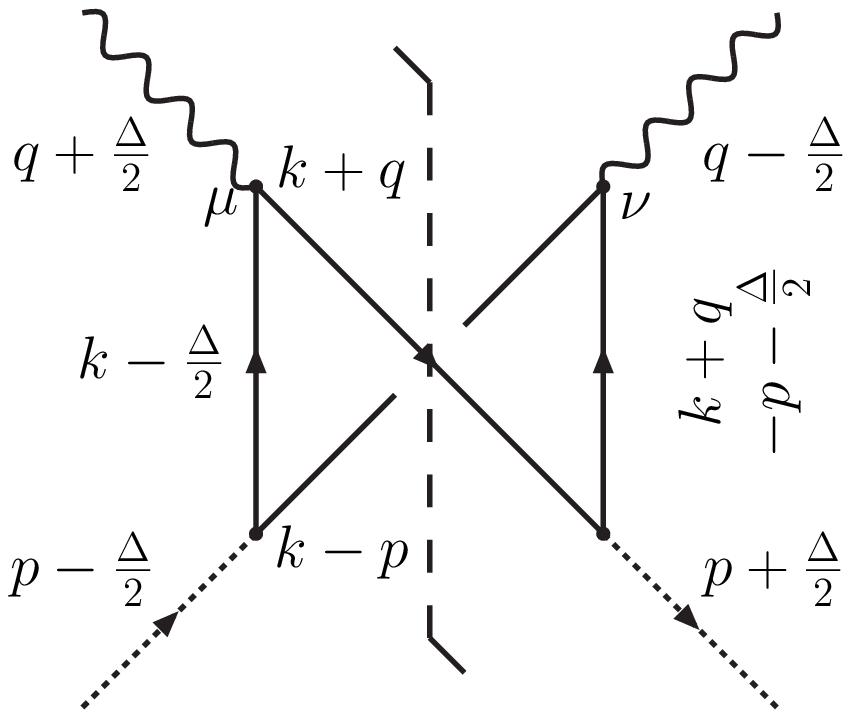}
}
\caption{Lowest order diagrams contributing to the imaginary part of the 
$\gamma^{\star} \pi \rightarrow \gamma^{\star} \pi$ scattering amplitude}
\label{fig:diagrams}
\end{figure}

The diagrams contributing to the imaginary part of the amplitude are displayed on
Fig.~\ref{fig:diagrams}. The Lorentz invariants are $t = \Delta^2$, $Q^2 = -q^2$,
 $x = Q^2/2p\cdot q$, $\xi = \Delta\cdot q / 2p\cdot q$.
The hadronic tensor reads
\begin{eqnarray}
\label{eq:tmunu}
&\textstyle{T_{\mu\nu} =
- {\cal P}_{\mu\sigma} g^{\sigma\tau} {\cal P}_{\tau\nu}
F_1
+ \frac{{\cal P}_{\mu\sigma} p^\sigma p^\tau {\cal P}_{\tau\nu}}{p \cdot q}
F_2
+ \frac{{\cal P}_{\mu\sigma} (p^\sigma (\Delta^\tau-2\xi p^\tau)
+  (\Delta^\sigma-2\xi p^\tau) {\cal P}_{\tau\nu}}{2p \cdot q}
F_3\nonumber}\\
&\textstyle{+ \frac{{\cal P}_{\mu\sigma}
(p^\sigma (Delta^\tau-2\xi p^tau)
-  (\Delta^\sigma-2\xi p^\sigma)p^\tau) {\cal P}_{\tau\nu}}{p p \cdot q}
F_4
+ {\cal P}_{\mu\sigma}
(\Delta^\sigma-2\xi p^\sigma) (\Delta^\tau-2\xi p^\tau) {\cal P}_{\tau\nu}
F_5.}
\end{eqnarray}
where $\cal P$ is the projector built on the metric tensor together with the momenta
of the ingoing and outgoing photons.
The Pion Quark coupling is described by the Lagrangian density
$ {\cal L}_{int}= i g (\overline{\psi} \vec\tau \gamma_5 \psi). \vec\pi$,
where $\psi$ stands for the quark field and $\vec{\pi}$ for the pion field
while $\vec{\tau}$ is the isospin operator.
The cut-off, which accounts for the finite size of the pion, is
imposed by requiring that the relative four-momentum squared of the quarks
inside the pion be smaller than $\Lambda^2$ at one of the quark-pion vertices of
each diagram \cite{ours2}.
One should note that imposing these conditions leads to a suppression
of the crossed diagrams by a factor $\Lambda^2/Q^2$ compared with
the box diagrams \cite{ours}.
In the following we keep the value of the coupling constant $g$ to its
diagonal case value, obtained by imposing the sum rule on $F_1$, that is
imposing that there be two valence quarks inside the pion.

\ \\

\section{Results for the structure functions}

Having obtained the imaginary parts of the five structure functions $F_i$
by projection of the amplitude on the corresponding tensors, we
look for their asymptotic behaviour at large $Q^2$, and find
relations between them at leading order. These relations read

\begin{eqnarray}
\label{eq:F_irel}
F_2&=2xF_1+{\cal O}(1/Q^2),&F_3=\frac{2x\xi}{\xi^2-1}F_1+{\cal O} (1/Q^2),\nonumber\\
 F_4&=\frac{2x}{\xi^2-1}F_1+{\cal O} (1/Q^2),&F_5={\cal O} (1/Q^2).
\end{eqnarray}

Getting such simple relations, similar to the Callan-Gross relation 
between the diagonal form factors, constitutes a remarkable result
of our model. We shall see that these {\it{Generalised Callan-Gross Relations}}
will lead to relations between the generalized parton distributions.

\section{Generalized parton distributions}

Following the twist-three tensor analysis of Ref. \cite{Bel} that links
the twist-two ${\cal H}$ and twist-three ${\cal H}^3, \tilde {\cal H}^3$ form factors
to the tensorial content of $\textstyle{T_{\mu\nu}}$,
we can relate the $F_i$ structure functions to the form factors \cite{ours2}.
The imaginary parts of the form factors yield the corresponding Generalized Parton Distributions
up to a constant factor.
Normalizing the structure functions by use of the same factor, we can
relate them to the GPD's, getting:

\begin{eqnarray}\label{eq:F15}
F_{1n}=H, 
\hbox{     }
F_{2n}=2x{H}, 
\hbox{     }
F_{3n}=\frac{2x}{x^2-\xi^2}\left({H}^3x^2+\tilde{H}^3\xi x-{H}\xi\right),\nonumber\\
F_{4n}=\frac{2x}{x^2-\xi^2}\left({H}^3\xi x+\tilde{H}^3x^2-{H}x\right),
\hbox{     }
F_{5n}={\cal O}(1/Q^2).
\end{eqnarray}

So that, with the help of our {\it{Generalised Callan-Gross Relations}}
we can write 
\begin{eqnarray}\label{eq:HH3rel}
\tilde{H}^3=\frac{(x-1)}{x(\xi^2-1)}H
\hbox{   and   }
{H}^3=\frac{(x-1)\xi}{x(\xi^2-1)}H = \xi\tilde{H}^3.
\end{eqnarray}

We compared relations (\ref{eq:HH3rel}) with the results of the Wandzura-Wilczek
approximation \cite{WW} applied to our results for $H$, and found these
to be very different.

Hence we conclude that our relations (\ref{eq:HH3rel}) are new, and do not correspond
 to kinematical twist corrections but come from the dynamics of the model we use,
including its finite-size content.


\begin{theacknowledgments}
  The authors thank M.V.Polyakov for useful comments.
  This work was performed with the help of the ESOP
  collaboration (European Union contract HPRN-CT-2000-00130)
\end{theacknowledgments}

\bibliographystyle{aipproc}   

\begin{thebibliography} {10}

\bibitem{off}
W.~Broniowski {it et al.},
 \emph{Phys. Lett. B}, \textbf{574}, 57 (2003) ;
S. Dalley, \emph{Phys. Lett. B} \textbf{570}, 191 (2003) [arXiv:hep-ph/0306121];
L.~Theussl {\it et al.},
 \emph{Eur. Phys. J. A}, \textbf{20}, 483 (2004) . 

\bibitem{ours}
F.~Bissey, J.R.~Cudell, J.~Cugnon, M.~Jaminon, J.P.~Lansberg and P.~Stassart,
 \emph{Phys. Lett. B}, \textbf{547}, 210 (2002) [arXiv:hep-ph/0207107];
J.P.~Lansberg, F.~Bissey, J.R.~Cudell, J.~Cugnon, M.~Jaminon and P.~Stassart, \emph{AIP Conf. Proc.} \textbf{660} 339 (2003)  [arXiv:hep-ph/0211450].

\bibitem {Bel}
A.~V.~Belitsky {\it et al.},
 \emph{Phys. Rev. D}, \textbf{64}, 116002 (2001) [arXiv:hep-ph/0011314].


\bibitem {ours2}
F.~Bissey, J.R.~Cudell, J.~Cugnon, J.P.~Lansberg and P.~Stassart,
 \emph{Phys. Lett. B}, \textbf{587}, 189 (2004).
\bibitem {WW}
S.~Wandzura and F.~Wilczek,
 \emph{Phys. Lett. B}, \textbf{ 72}, 195 (1977) .
\end{thebibliography}

\end{document}